\documentclass[10pt, a4paper]{article}
\usepackage {fullpage}						
\usepackage {ucs}
\usepackage [utf8x]{inputenc}
\usepackage {graphicx}
\usepackage [bookmarks=false, pdfstartview={FitH}, colorlinks=true, linkcolor=blue, citecolor=blue, urlcolor=blue]{hyperref} 
\usepackage {amsfonts,amsmath,amsthm,amssymb}
\usepackage {latexsym}
\usepackage {graphics, graphicx}
\usepackage {fancyhdr}
\usepackage[table]{xcolor}
\usepackage {epstopdf} 						
\usepackage {multicol}						
\usepackage {multirow}  					
\usepackage {soul} 							
\usepackage {authblk} 						
\usepackage {datetime} 						
\usepackage {subfigure}						
\usepackage {parskip}						
\usepackage {setspace}						
\usepackage[font=small,textfont=it,width=0.97\textwidth]{caption}		
\usepackage{mathtools}						

\newtheorem{remark}{Remark}

\renewcommand{\(}{\left(}
\renewcommand{\)}{\right)}

\newcommand{\go}{[g]_0}
\newcommand{\gb}{[g]_b}
\newcommand{\gf}{[g]_f}

\newcommand{\ro}{[r]_0}
\newcommand{\rb}{[r]_b}
\newcommand{\rf}{[r]_f}

\newcommand{\cc}{c^\text{CSC}}
\newcommand{\cd}{c^\text{DCC}}
\newcommand{\lc}{\lambda^\text{CSC}}
\newcommand{\ld}{\lambda^\text{DCC}}

\newcommand{\memt}{\mu_\text{EMT}}

\newcommand {\startenv} {\vskip 0.05em
\begin{tabular}{||l}\parbox[t]{0.95\linewidth}}
\newcommand {\stopenv} {\end{tabular}\vskip 0.05em}

\def\gray#1{{\color{gray}\text{#1}}}

\def\comm#1{{\color{red}\ul{#1}}}
\newcommand{\change}[2]{{}{#2}}

\title{A mathematical insight in the epithelial–mesenchymal-like transition in cancer cells and its effect in the invasion of the extracellular matrix}

\author{
	Nadja Hellmann\thanks{Institute of Molecular Biophysics, University of Mainz, Germany \hfill {\tt nhellman@uni-mainz.de}}~{},
	Niklas Kolbe\thanks{Institute of Mathematics, University of Mainz, Germany\hfill {\tt kolbe@uni-mainz.de}}~{},
	Nikolaos Sfakianakis\thanks{Institute of Mathematics, University of Mainz, Germany \hfill {\tt sfakiana@uni-mainz.de}}~{}
	}
\date{}

\begin{document}
\maketitle


\begin{abstract}
	Current biological knowledge supports the existence of a secondary group of cancer cells within the body of the tumour that exhibits stem cell-like properties. These cells are termed \textit{cancer stem cells}, and as opposed to the more usual \textit{differentiated cancer cells}, they exhibit higher motility, they are more resilient to therapy, and are able to metastasize to secondary locations within the organism and produce new tumours. The origin of the \textit{cancer stem cells} is not completely clear; they seem to stem from the \change{more usual}{} \textit{differentiated cancer cells} via a transition process related to the \textit{epithelial-mesenchymal transition} that can also be found in normal tissue.
	
	In the current work we \change{focus on the modelling and the numerical study of}{model and numerically study} the transition between these two types of cancer cells, and the resulting ``ensemble'' invasion of the \textit{extracellular matrix}. \change{}{This leads to the derivation and numerical simulation of two systems: an algebraic-elliptic system for the transition and an advection-reaction-diffusion system of Keller-Segel taxis type for the invasion.}
\end{abstract}


	Cancer research aims to understand the causes of cancer and to develop strategies for its diagnosis and treatment. In this effort, mathematics contributes with the modelling of the biological processes, and the corresponding analysis and numerical simulations. As a research field, it has been very active since the 1950s; see e.g. \cite{Nordling.1953, Armitage.1954, Fisher.1958} and has spanned over a wide range of applications from intracellular bio-chemical reactions to cancer growth and metastasis, see e.g. \cite{Preziosi.2003,Perumpanani.1996, Anderson.2000, Gerisch.2008, Ganguly.2006,Michor.2008, Czochra.2012,7,Johnston.2010,Vainstein.2012, Gupta.2009}. 	
	
	Intrinsically, cancer cells exhibit higher proliferation rates than normal cells. Recent evidence though have revealed the existence of a subpopulation of cancer cells that posses lower proliferation rates than the rest of cancer cells, exhibits stem-like properties such as self-renewal and cell differentiation, and are able to metastasize i.e. to detach from the primary tumour, invade the vascular system, and afflict secondary sites \cite{Brabletz.2005, Mani.2008}. These are termed \textit{Cancer Stem Cells} (CSCs) and constitute a small part of the tumour, while the bulk of the tumour consists of the more proliferative \textit{Differentiated Cancer Cells} (DCCs), \cite{Thiery.2002, Reya.2001} 
			
	The origin of CSCs is not completely known; a current theory states that they are \textit{de-differentiated} cancer cells originating from the more usual DCCs \cite{Gupta.2009, Reya.2001}. This type of transition of cancer cells, influences their \textit{cellular potency} 
	and seems to be related to a type of cellular differentiation program that can be found also in normal tissue, the \textit{Epithelial-Mesenchymal Transition} 
	(EMT) \cite{Mani.2008}.  The EMT takes place in the first step of metastasis, the invasion of the \textit{Extracellular Matrix} 
	(ECM) by the primary tumour, cf. \cite{Thiery.2002,Katsumo.2013}.
	

	The EMT is triggered by the micro-environment of the cell \cite{Radisky.2005}. The protein\change{s}{}, for example, \textit{Epidermal Growth Factor} (EGF) \change{and \textit{Transforming Growth Factor beta} (TGF-$\beta$) and their}{and its} corresponding cellular receptors EGFR \change{s and TGFBRs} plays pivotal role in this triggering \cite{3,Shien}. \change{Briefly, in the case of EGFs, this}{This } process can be \change{}{briefly} described as follows: the EGFs function as ``on''/``off'' switches that stimulate cellular growth and proliferation. In certain pathological mutations, they get stuck in the ``on'' position and cause unregulated cell growth, untimely EMT, and over-expression of EGFRs. The over-expression though, of EGFRs has been connected to the tumor formation in mammary epithelial cells, see e.g. \cite{2}. 


	From mathematical perspective, the modelling of ECM invasion takes into account a special type of taxis, namely \textit{haptotaxis}. \change{The availability and the gradient of adhesion sites on the ECM or of ECM-bound chemoattractants, drives the cells to a biased directional movement.}{In haptotaxis, the cells move biased, along directions being dictated by the availability or the gradients of either extracellular adhesion sites (which are located on the ECM), or of ECM-bound chemoattractants.}
	
	From the macroscopic point of view \change{chemo-hatotaxis}{haptotaxis (as the other forms of taxis)} is typically modelled under the paradigm of Keller-Segel systems which was first introduced in \cite{KS.1970} to describe the aggregations of \textit{slime mold}. 
	Since their first derivation, they have been successfully applied in a wide range of biological phenomena spanning from bacterial aggregations to wound healing, and cancer modeling. In cancer growth modelling in particular, Keller-Segel type systems were extended to include also enzymatic interactions/reactions, yielding \textit{Advection-Reaction-Diffusion} (ARD) models, see for instance the generic haptotaxis system proposed in \cite{Anderson.2000} (promoting ECM degradation by \textit{Matrix Metalloproteinases} 
	(MMPs)), or the chemo-hapto-taxis system proposed in \cite{Chaplain.2005} that promotes the role of the enzyme \textit{plasmin}. 
	
	From a numerical point of view, the treatment of this type of problems is a challenging task, since their solutions develop (generically) heterogeneous spatio-temporal dynamics in the form of merging/emerging clusters of cancer cell concentrations, see for instance Figure \ref{fig:invasion}. To treat these dynamics we employ a second order Finite Volume method equipped combined with a third order Implicit-Explicit time integration scheme, see Section \ref{sec:invasion} and our previous works \cite{Sfakianakis.2014b,Kolbe.2013} for more details.

	In this work, our objective is twofold; first we model the EMT transition from DCCs to CSCs\change{ which is}{,} triggered by the availability of \change{EGF}{free EGF molecules} in the extracellular environment, and of the corresponding \change{}{EGFR} cellular receptors, and secondly we embed this EMT model in an ECM invasion system that addresses the ``ensemble'' CSCs and DCCs. We subsequently perform \change{}{some} numerical \change{simulations}{experiments} to investigate \change{some of the}{} properties of \change{these two systems}{the combined system}. 

	In more detail, we construct a parabolic system describing the concentrations of EGFs and EGFRs and their attachment/detachment dynamics. These processes are assumed to be much faster than the cellular motility and invasion of the ECM, hence we rescale the previously developed system \change{to}{and} deduce an elliptic system that addresses \change{the concentrations of}{} the free and bound EGF\change{ molecules}{}. The rescaled elliptic system serves as the large time asymptotic limit of the parabolic system. \change{We subsequently make use of the \comm{(citation)} and deduce the coefficient of the EMT for the ECM invasion system.} 

	To model the ECM invasion, we employ an ARD haptotaxis model that we adjust to include the CSC-DCC ``ensemble''\change{}{ and the EMT between them}. In particular we assume that the cancer cells diffuse in the environment, move responsively to the gradients of the ECM, proliferate in a logistic manner, and most notably undergo EMT. \change{In addition, we}{We also} include the ECM degradation by a matrix degenerating MMP, which in turn is assumed to be produced by both families of cancer cells.

	With \change{}{the combination of} these models we are able to reproduce the \change{}{current} biomedical understanding that the CSCs a) stem from the bulk of DCCs, b) they separate themselves from the main body of the tumour, c) they invade the ECM while exhibiting highly dynamic \change{concentration}{clustering} phenomena.

	\change{}{Similar ECM invasion systems that can be found in the literature, take into account multiple subpopulations of cancer, can be found (see e.g. \cite{Andasari.2011,Domschke.2014}). These systems however control the transition between the subpopulations in terms of time dependent Heavyside step functions, while in our approach the suggested models connect the microscopic dynamics of EGF with the macroscopic dynamics of EMT and ECM invasion.}


\section{The EMT coefficient}
\change{}{
	Typical modeling of the ECM invasion by a single type of cancer cell includes terms for diffusion, taxis, reaction, and proliferation. For two types of cancer cells, the corresponding equations apply with an addition of a coupling term, which in our case describes the transition from DCC to CSC via EMT, i.e. 
	\begin{equation}\label{eq:generic}
	\begin{aligned}
		\frac{\partial \cc}{\partial t}&=\gray{diffusion}+\gray{taxis} + \memt \;\cd
												+\gray{reaction}+\gray{proliferation}\\ 
		\frac{\partial \cd}{\partial t}&=\gray{diffusion}+\gray{taxis} - \memt \;\cd
												+\gray{reaction}+\gray{proliferation}
	\end{aligned}.\end{equation}
	where $\cc$, $\cd$ represent the densities of the CSCs and the DCCs respectively. The coefficient $\memt$ describing the probability of the EMT  transition is given by Michaelis-Menten kinetics (see also \cite{Zhu.2011})
	\begin{equation}\label{eq:muemt}
		\memt=\mu_0 \frac{\gb^\text{DCC}}{\mu_{1/2} + \gb^\text{DCC}},
	\end{equation}
	where $\gb^\text{DCC}$ represents the concentration of the EGF molecules that are bound on the corresponding EGFR receptors on the membrane of the DCC cells, and where $\mu_0$ is the maximum EMT rate, and $\mu_{1/2}$ is the Michaelis constant, i.e. the bound EGF needed to generate half maximal EMT. 
}

	In addition to $\gb^\text{DCC}$, we set $\gf$ to represent the concentration of free EGF molecules (in the extracellular environment), $\go$ the total (free and bound on both DCCs and CSCs) concentration of EGF. Accordingly, $\ro$, $\rb$, and $\rf$ \change{will be}{denote} the total, occupied, and free receptors on the surface of both types of cancer cells. 
	
	Throughout this work we assume that all $[g]_\text{\textvisiblespace}$ and $[r]_\text{\textvisiblespace}$ concentrations depend on time and on space\change{}{; this is so} since the free EGF molecules are able to translocate (for example diffuse) in the extracellular environment, and the bound EGFs and the EGFRs are to be found on the surfaces of the (also time and space dependent) CSCs and DCCs. We assume moreover that the total EGFs and EGFRs obey local mass conservation properties, i.e.
	\begin{align}
		\go(t,x)&=\gf(t,x) + \gb(t,x) \label{eq:mass_EGF}\\
		\ro(t,x)&=\rf(t,x) + \rb(t,x) \label{eq:mass_EGFR}
	\end{align}
	for every $t\geq0$ and $x\in \Omega\subset\mathbb{R}$ bounded.
	
\paragraph{The parabolic system.}
	The dynamics dictate that free EGF molecules bind onto free EGFR receptors with rate $k_+$, while at the same time bound EGF molecules detach with rate $k_-$ (see also \cite{Zhu.2011}). It is assumed that the $k_+$ and $k_-$ rates are the same for both types of cancer cells. \change{Hence}{That is,} the time evolution of both the free and the bound EGF molecules is given by the  system:
	\begin{equation}\label{eq:parab}	
		\left\{ \begin{aligned}
           \partial_\tau \gb &= && k_+ \gf \rf - k_- \gb, \\
           \partial_\tau \gf &= D_f \Delta \gf - && k_+ \gf \rf  + k_- \gb.
          \end{aligned}
		\right.
	\end{equation}
	where $\tau$ is the corresponding time variable\change{ and where we have also assumed}{. We have also assumed} that the free EGF molecules diffuse in the extracellular environment.
	
	Recalling now the mass conservation of the EGFRs \eqref{eq:mass_EGFR}, we \change{see}{can deduce} that:
	\begin{align}
		\rf & = \ro - \rb \nonumber \\
			& = \lc \cc + \ld \cd - \gb,
	\end{align}
	where $\lc, \ld$ represent the total ``number'' of EGFR receptors per cell per family, and where we have used that the bound EGF molecules coincide with the bound EGFR receptors. 
	
\paragraph{The elliptic system.}
	\change{We also assume that the}{The} EGF dynamics (i.e attachment/detachment  onto EGFRs and diffusion) \change{is}{are} much faster than the dynamics of EMT and ECM invasion. We rescale hence the time variable \change{}{$\tau$} in System \eqref{eq:parab}, to match the time scale $t$ of the systems \eqref{eq:generic}, \eqref{eq:invasion}.   To this end, we first rewrite \eqref{eq:parab} in the form 
	$$\partial_\tau \b u = D\Delta \b u + p(\b u),$$
	where $u=(\gb,~\gf)^T, ~ D = (0,~D_f)^T$, and where $p$ denotes the nonlinear reaction terms in \eqref{eq:parab}. We employ \change{}{now} the change of variables 
	\begin{equation}\label{eq:ch_var}
		\b v(t,x) = \b u(\tau(t),x),
	\end{equation} 
	\change{where}{with} the ``much faster'' time variable $\tau$ \change{is}{} given by
	\begin{equation}
		\tau(t) =  \frac{t}{\varepsilon}, \quad 0<\varepsilon <<1.
	\end{equation}
	Relation \eqref{eq:ch_var} yields after $\frac{\partial}{\partial t}$ differentiation as
	\begin{equation} \label{eq:with_eps}
		\varepsilon \partial_t \b v = D\Delta \b v + p(\b v),
	\end{equation}
	since\change{, after \eqref{eq:ch_var},}{}
	$$\Delta \b v(t,x)= \Delta \b u(\tau(t),x),\quad p(\b v(t,x)) =  p(\b u(\tau(t),x)),\quad \frac{\partial \tau(t)}{\partial t} = \frac 1\varepsilon, \quad \frac{\partial x }{\partial t}=0.$$
	We denote the above introduced variables in the $t$ timescale as follows
	$$\begin{pmatrix}g_b (t,x) \\ g_f(t,x) \end{pmatrix} = \begin{pmatrix}\gb(\tau(t),x) \\ \gf(\tau(t),x) \end{pmatrix} = \b v(t,x).$$ 

	Taking now in \eqref{eq:with_eps} the formal limit as $\varepsilon\rightarrow 0$, we deduce the elliptic system that corresponds to \eqref{eq:parab}:
	\begin{equation*}
		\left\{ \begin{aligned}
           0&= && k_+ g_f (\lc\cc +\ld\cd - g_b) - k_- g_b, \\
           0 &= D_f \Delta g_f - && k_+ g_f (\lc\cc +\ld\cd - g_b)  + k_- g_b.
          \end{aligned} \right.
	\end{equation*}
	or
	\begin{equation}\label{eq:ellip}
		\left\{ \begin{aligned}
           0 &= \frac{1}{k_D} g_f (\lc\cc +\ld\cd - g_b) - g_b, \\
           0 &= \Delta g_f.
          \end{aligned} \right.
	\end{equation}
	where $k_D=\frac{k_-}{k_+}$ represents the attachment/detachment ratio of the EGFs onto the EGFRs. Though \eqref{eq:ellip} does not include time derivatives, both concentrations $g_b$ and $g_f$ depend on time $t$ through the densities of the cancer cells $\cc$ and $\cd$.

\paragraph{Neumann and additional conditions}
	We augment the system \eqref{eq:ellip} with homogeneous Neumann boundary conditions, 
	\begin{equation}\label{eq:neumann}
		\frac{\partial g_f(x)}{\partial \bf n} = 0,\quad x\in\partial \Omega.
	\end{equation}
	\begin{remark}
		The system \eqref{eq:ellip}-\eqref{eq:neumann} has no unique solution. This can be seen by making the ansatz $g_b = K(\lc\cc +\ld\cd), ~K\in(0,1)$. For this choice the constant function
	 	$$g_f = k_D\frac{K}{1-K},$$
 		satisfies the algebraic equation of \eqref{eq:ellip}, the boundary condition \eqref{eq:neumann}, and $\Delta g_f = 0$. Thus for every $K$, a classical solution of \eqref{eq:ellip}, \eqref{eq:neumann} exists.
	\end{remark}

	To recover uniqueness in the system \eqref{eq:ellip}-\eqref{eq:neumann} we impose an additional condition, namely we set the total mass of EGF to be constant (in time $t$), 
	\begin{equation}\label{eq:egf_mass}
		M = \int_\Omega g_0(t,x)\,dx=\int_\Omega g_b(t,x)\,dx + \int_\Omega g_f(t,x)\, dx.
	\end{equation}
	Condition \eqref{eq:egf_mass} comes as a natural consequence of the system \eqref{eq:parab} endowed with zero Neumann boundary conditions. The total mass of EGF in this system does not change with time $\tau$ since, after integration of the second equation of \eqref{eq:parab} over $\Omega$, we see that
	$$\frac{d}{d\tau} \int_\Omega \gb \, dx +\frac{d}{d\tau} \int_\Omega \gf \, dx = \int_\Omega \partial_\tau \gb \, dx + \int_\Omega -\partial_\tau \gb \,  +  D_f \Delta \gf\, dx = 0.$$
	In effect, the total mass of EGF in the elliptic system \eqref{eq:ellip} given by \eqref{eq:egf_mass} will be equal to 
	$$M =  \int_\Omega \gf^0\, dx + \int_\Omega \gb^0\, dx,$$
	in the parabolic system \eqref{eq:parab} for given initial concentrations $\gb^0$ and $\gf^0$.
	
\paragraph{Solution of the elliptic system.}
	\begin{figure}[t]
		\centering
		\includegraphics[width=0.9\textwidth]{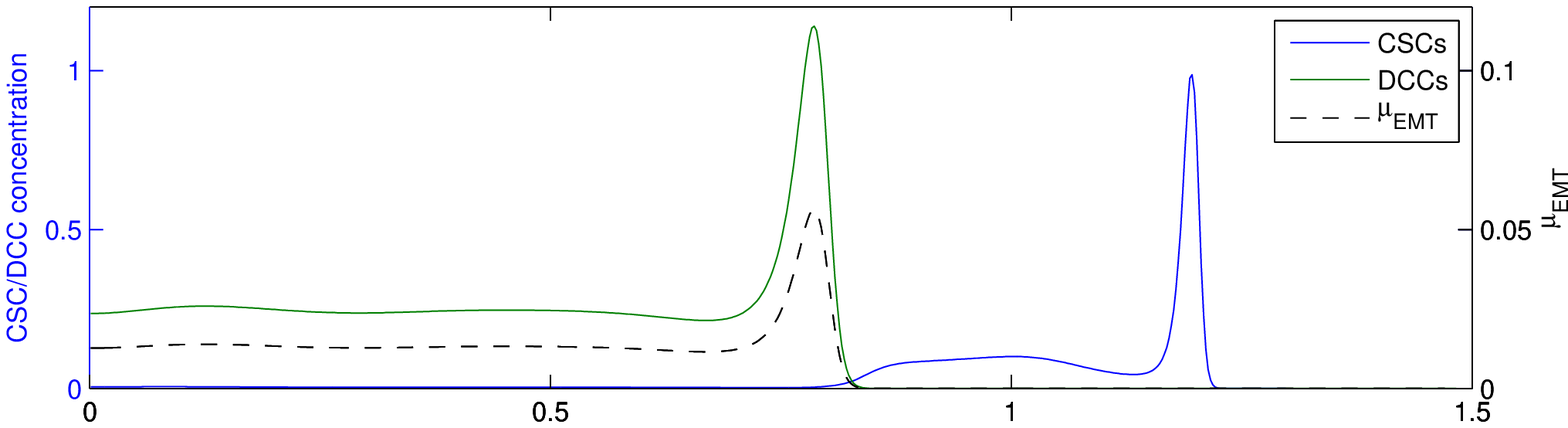}
		\caption{Plot of $\memt$ for the system \eqref{eq:invasion}-\eqref{eq:params_2} at time $t=6$. Note \change{it's space dependence in}{that it is space dependent and that it ``lives'' over} the support of the DCCs.}\label{fig:memt}
	\end{figure}
	
	We note at first that \change{in the system}{the Laplace equation in System} \eqref{eq:ellip}\change{, the Laplace equation on the regular domain $\Omega$}{} has a classical solution for which the gradient $\nabla g_f$ is unique. We can thus deduce that $g_f=\tilde g_f$ is constant. Accordingly, the first equation of the system \eqref{eq:ellip} implies 
	\begin{align}
		g_b &= \frac{\frac{1}{k_D}\tilde g_f}{ 1 + \frac{1}{k_D}\tilde g_f} \(\lc\cc +\ld\cd\),\nonumber \\
			&= \frac{\tilde g_f}{ k_D + \tilde g_f} \(\lc\cc +\ld\cd\).\label{eq:g_b_full}
	\end{align}
	Similar, but slightly more involved, is the derivation for the DCC-bound EGF. We repeat the assumption that the attachment/detachment rates $k_+$, $k_-$ are the same for both families of cancer cells, and that the superposition principle $[r]_f=[r]_f^\text{CSC}+[r]_f^\text{DCC}$ with $[r]_f^\text{CSC}$, $[r]_f^\text{DCC}$ the free EGFRs on the CSCs, DCCs respectively, holds. Accordingly the DCC-bound EGF is given by
	\begin{equation}\label{eq:g_b}
		g_b^\text{DCC}  = \frac{\tilde g_f}{k_D + \tilde g_f} \ld\cd.
	\end{equation}
	
	On the other hand, the free EGF $\tilde g_f$ (which is constant) satisfies the quadratic equation
	\begin{equation}\label{quadratic_K}
		0 = |\Omega|\, \frac{1}{k_D} \tilde g_f^2 + \( |\Omega| + \frac{1}{k_D} \int_\Omega \lc\cc +\ld\cd dx - M \frac{1}{k_D} \)\tilde g_f - M, 
	\end{equation}
	obtained after integration of \eqref{eq:ellip} with respect to $x$ and taking \eqref{eq:egf_mass} into account. \change{The existence of a single positive root of}{The equation} \eqref{quadratic_K} \change{stems from the positivity of the discriminant}{has a single positive root since,}
	\begin{equation}\label{eq:discr}
		\Delta = \left( |\Omega| + \frac{1}{k_D}\int_\Omega \lc\cc +\ld\cd dx - M \frac{1}{k_D} \right)^2 + 4\frac{1}{k_D}\,|\Omega| M\geq 0
	\end{equation}
	and \change{the negativity of the term}{}
	$$ -\frac{k_D M}{|\Omega|}\leq 0.$$ 
	\change{Hence, the unique admissible solution of}{The root itself i.e.} the free EGF, is given by
	\begin{align}
		\tilde g_f &= \frac{-\left( |\Omega| + \frac{1}{k_D}\, \int_\Omega \lc\cc +\ld\cd dx - M \frac{1}{k_D} \right) + \sqrt{\Delta}}{2 |\Omega|\, \frac{1}{k_D}}\nonumber\\
		&= \frac{-\left( |\Omega|k_D + \int_\Omega \lc\cc +\ld\cd dx - M \right) + k_D\sqrt{\Delta}}{2 |\Omega|}\label{explicit_K}
	\end{align}
	\begin{remark}
		We note that $\tilde g_f$, though constant in space, is not constant in time since $\int_\Omega \lc\cc +\ld\cd dx$ varies with time $t$. \change{Similarly}{In effect}, $g_b^\text{DCC}$ and $\memt$ depend on time. Note also that although $\tilde g_f$ is constant in space, $g_b^\text{DCC}$ \eqref{eq:g_b} is not constant in space\change{. In effect}{, so } neither is $\memt$ \eqref{eq:muemt} constant in space.
	\end{remark}
	Combining now \eqref{eq:muemt} with \eqref{eq:g_b} we deduce that
	\begin{equation}\label{eq:muemt_2}
		\memt=\mu_0\frac{\tilde g_f\ld\cd}{\mu_{1/2}k_D + \mu_{1/2}\tilde g_f + \tilde g_f\ld\cd},
	\end{equation}
	which is non constant in space.  \change{See}{We refer to} Figure \ref{fig:memt} for a graphical representation of $\memt$ in the experimental scenario \eqref{eq:invasion}-\eqref{eq:params_2} at time $t=6$.

\paragraph{Comparison of the systems \eqref{eq:parab} and \eqref{eq:ellip}.}
	\begin{figure}[t]
		\centering
		\subfigure[$t=0$]{
			\includegraphics[width=0.3\textwidth]{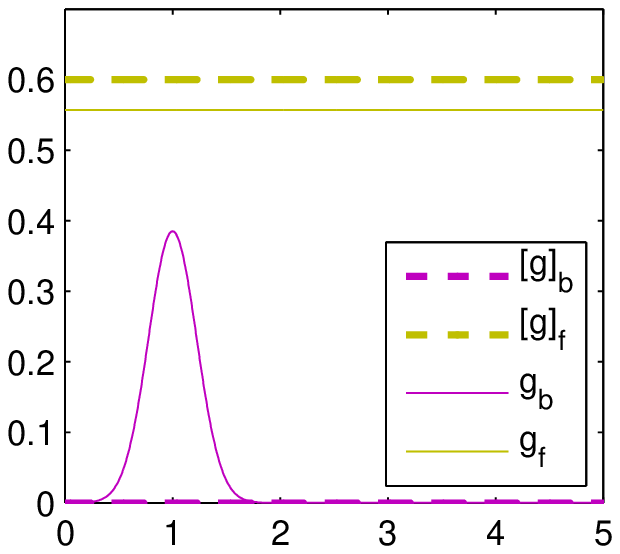}}
		\subfigure[$t=2$]{
			\includegraphics[width=0.3\textwidth]{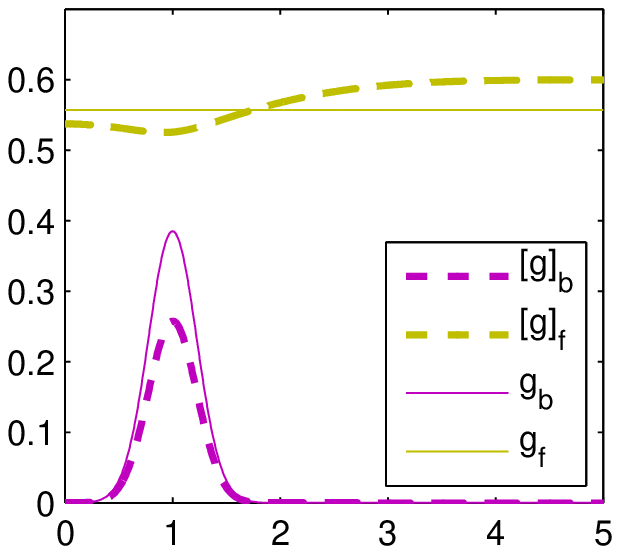}}
		\subfigure[$t=40$]{
			\includegraphics[width=0.3\textwidth]{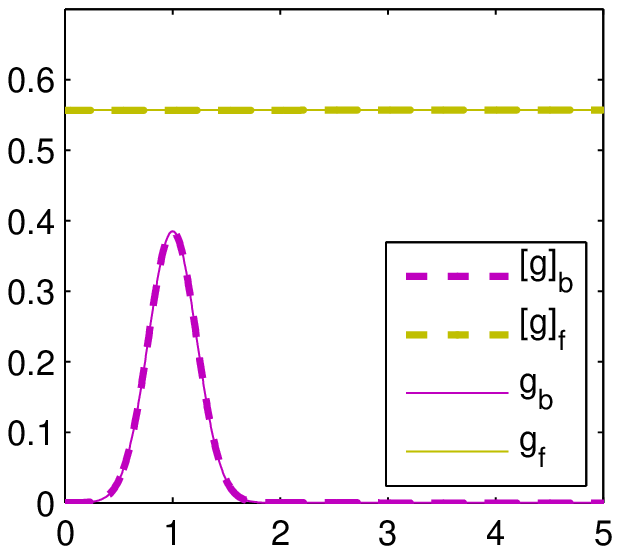}}
		\caption{Numerical solutions of \eqref{eq:parab} converge towards the numerical solution of \eqref{eq:ellip} as $t$ increases, if Neumann boundary conditions \eqref{eq:neumann} are assumed for both systems and if system \eqref{eq:ellip} is augmented by the condition $\int_\Omega g_b + g_f \, dx =  \int_\Omega \gf^0 + \gb^0\,  dx$. The parameters $D_f=0.4,~k_+ = 0.5,~k_-= 0.3$ and time independent $\lc\cc+ \ld\cd = e^{-10(x-1)^2}$ for $x\in \Omega=(0,5)$ were used.}\label{fig:neumann_prescribed_mass}
	\end{figure}
	
	The benefit that stems from the lack of \change{the} time variable in \eqref{eq:ellip} is the instantaneous ``sensing'' of the cancer cell concentration as opposed to the ``large'' time asymptotic convergence of the system \eqref{eq:ellip}. 
	
	The solutions of the elliptic system \eqref{eq:ellip}, coincide with the large time asymptotic solution of the parabolic system \eqref{eq:parab} endowed with zero Neumann boundary conditions \eqref{eq:neumann} if a selection criterion --in particular a prescribed total mass of EGF $M =  \int_\Omega \gf^0\, dx + \int_\Omega \gb^0\, dx$ in \eqref{eq:ellip}-- is employed. This can be also observed numerically in Figure \ref{fig:neumann_prescribed_mass}.

	Hence, at every instance of the ``large'' time scale $t$, at which EMT and cell movement take place, we can employ the elliptic system \eqref{eq:ellip} to compute the density \change{of the bound and of the total EGF and accordingly deduce the EMT coefficient \eqref{eq:muemt}.}{of the free EGF \eqref{explicit_K} and accordingly deduce the EMT coefficient from \eqref{eq:muemt_2}.}
	
	\change{From a numerical point of view, no expensive PDE solvers are needed to solve the elliptic system \eqref{eq:ellip} at every time step, since we know the exact solution \eqref{eq:g_b}, \eqref{explicit_K}.\comm{(expand, clarify, put in the intro)}}{From a numerical point of view, this process is cheap since no PDE solvers are needed.}	
\section{Invasion system}\label{sec:invasion}
	\begin{figure}[t]
		\centering
		\subfigure[$t=0.5$]{
			\includegraphics[width=0.3\textwidth]{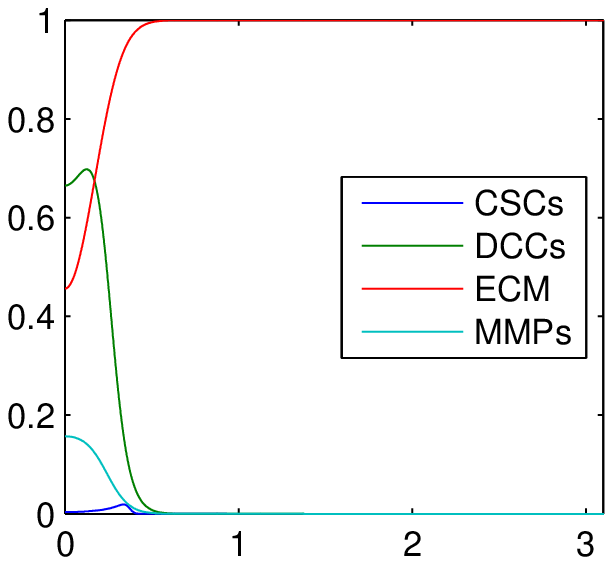}}
		\subfigure[$t=1.6$]{
			\includegraphics[width=0.3\textwidth]{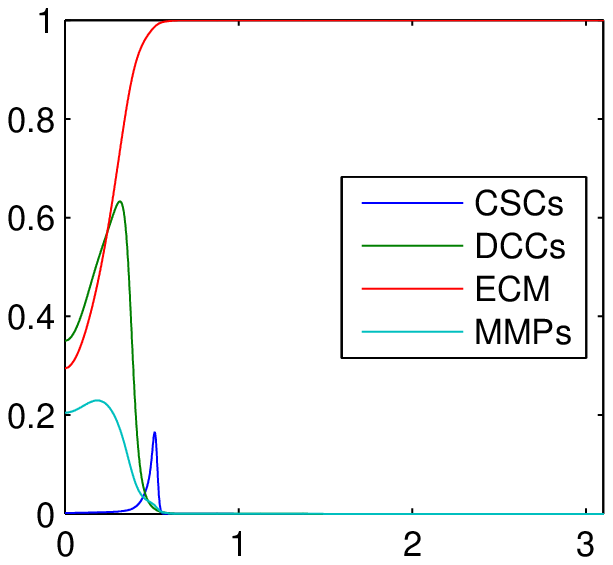}}
		\subfigure[$t=6.0$]{
			\includegraphics[width=0.3\textwidth]{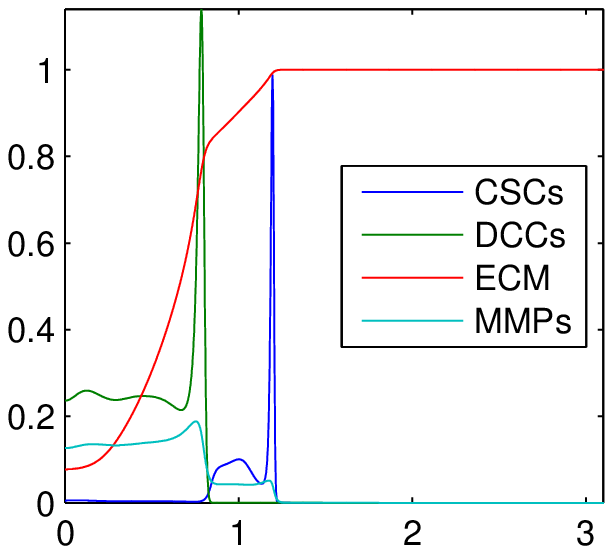}}\\
		\subfigure[$t=10.7$]{
			\includegraphics[width=0.3\textwidth]{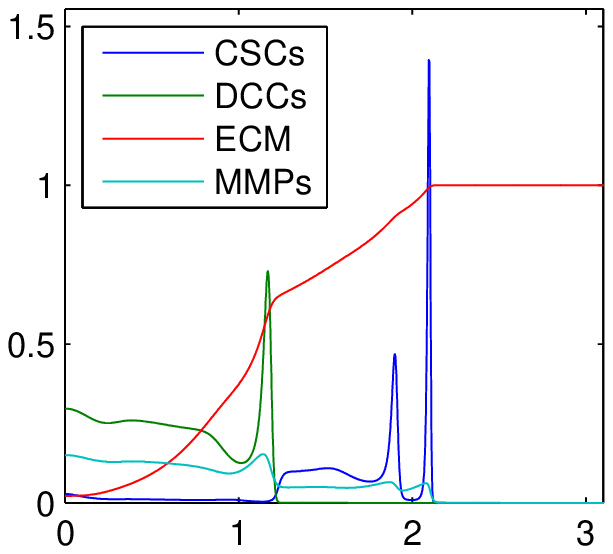}}
		\subfigure[$t=13.7$]{
			\includegraphics[width=0.3\textwidth]{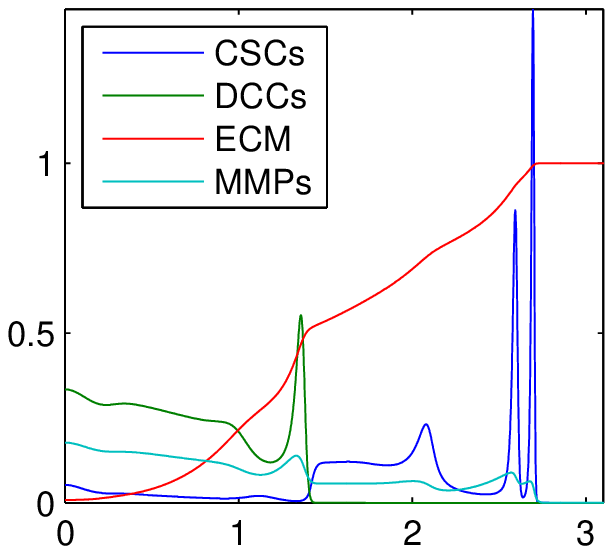}}
		\subfigure[$t=15.0$]{
			\includegraphics[width=0.3\textwidth]{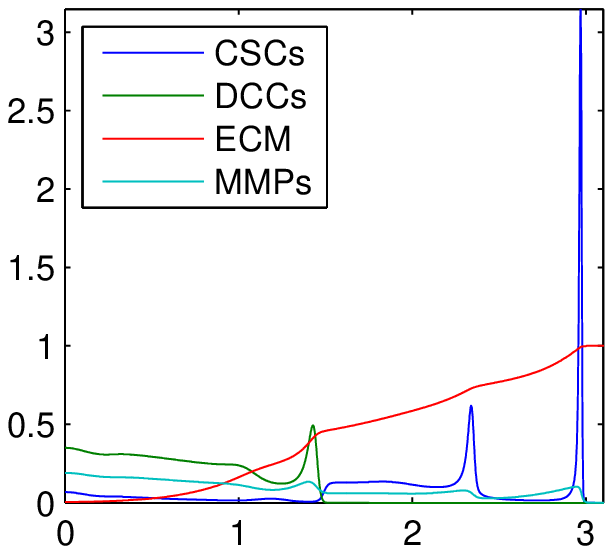}}
		\caption{Time evolution of the cancer invasion system \eqref{eq:invasion}-\eqref{eq:params_2}. (a) the CSCs are produced --via EMT-- by the DCCs,  (b)-(c) due to their higher motility, the CSCs escape the main body of the tumour and invade the ECM faster than the DCCs. Clearly two propagation fronts have been formed, (d)-(f) the CSCs present dynamic merging and emerging concentrations; the DCCs propagate into the ECM slower, smoother, and without the wealth of phenomena of the CSC. \change{\comm{Note that the ECM does not degrade where no cancer cells exixt.}}{}
		}\label{fig:invasion}
	\end{figure}
	
	\begin{figure}[t]
		\centering
			\includegraphics[width=0.45\textwidth]{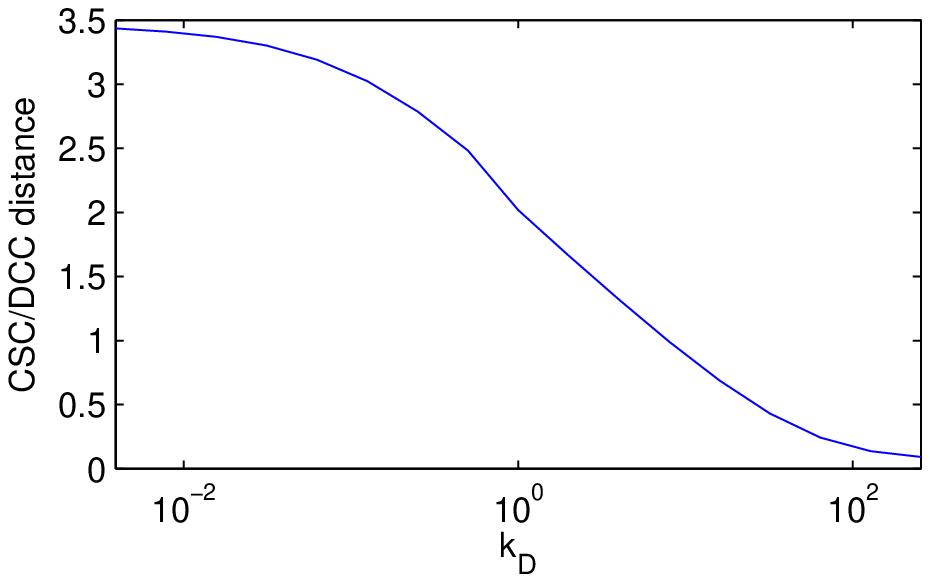}
			\includegraphics[width=0.45\textwidth]{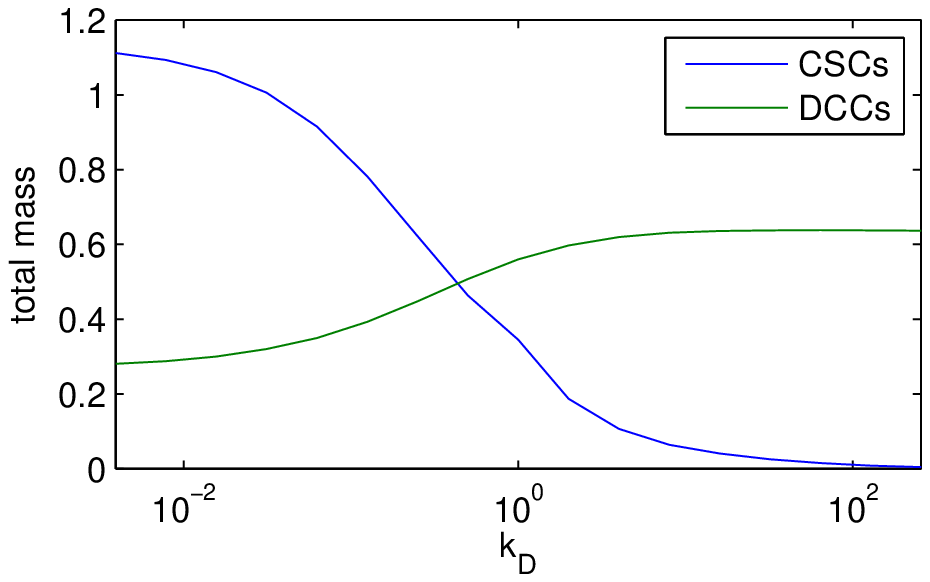} \\
				\includegraphics[width=0.45\textwidth]{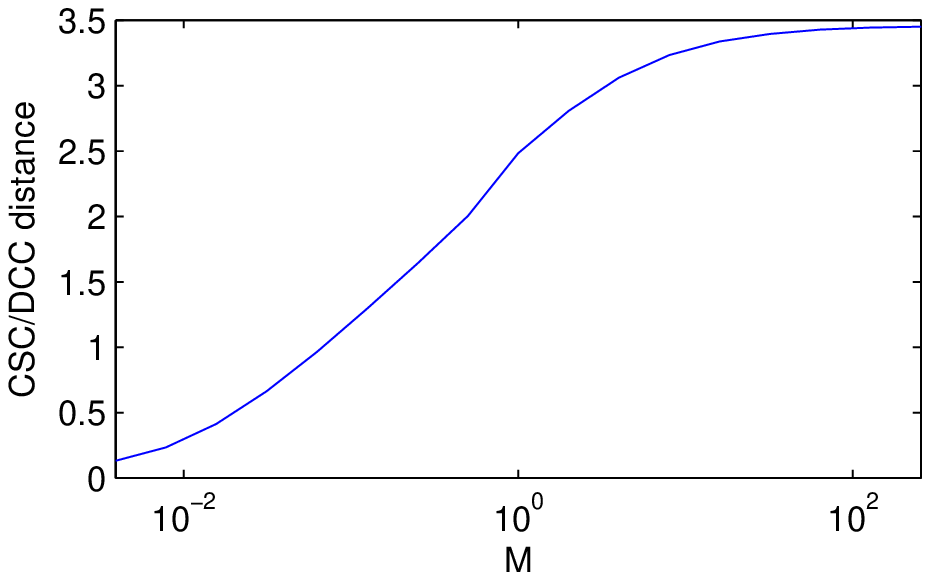} 
				\includegraphics[width=0.45\textwidth]{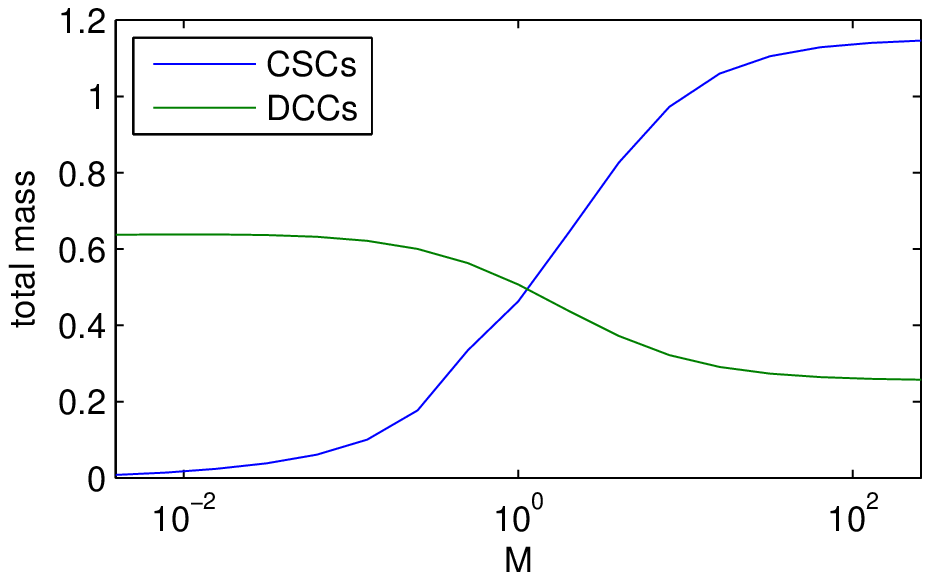} \\
				\includegraphics[width=0.45\textwidth]{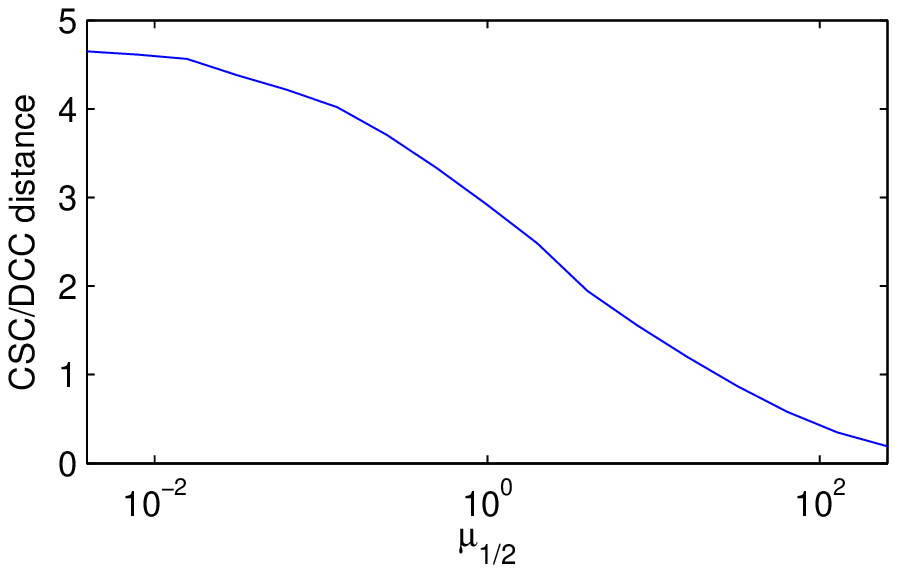}
				\includegraphics[width=0.45\textwidth]{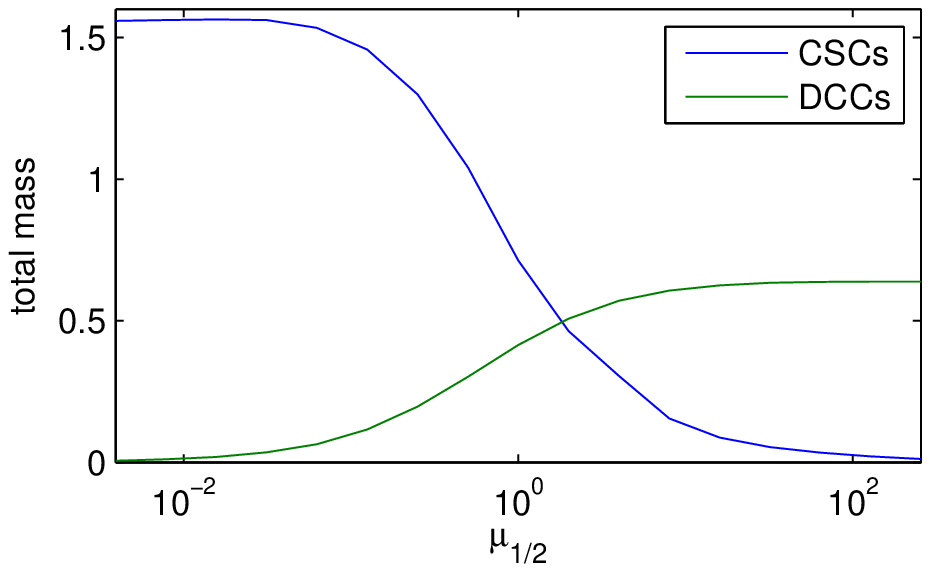}
		\caption{Propagating fronts-distance (left), and total masses (right) of the CSCs and the DCCs as functions  of $k_D=\frac{k_-}{k_+}$ (upper)\change{.}{, $M$ (middle), and $\mu_{1/2}$ (lower) in the case of the cancer invasion system \eqref{eq:invasion}-\eqref{eq:ic} with the rest EMT parameters as in $\mathcal{P}_{\memt}$, \eqref{eq:params_2}; the ECM parameters are as in $\mathcal P$, \eqref{eq:params}. An increase of both, the ratio $k_D$ and the Michaelis constant $\mu_{1/2}$ weakens the migration and the growth of CSCs. Increasing the EMT mass $M$ on the other hand yields a larger CSC/DCC distance and a higher amount of CSCs.}}\label{fig:parameter_dependence}
	\end{figure}
	
%
		
	The ECM invasion model we propose is based on the ``generic'' invasion model of \cite{Anderson.2000}. In this respect it includes haptotaxis, degradation of the ECM by MMPs, and production of the MMPs by the cancer cells. In addition to this basic model, we assume a secondary family of cancer cells, EMT transition between DCCs and CSCs, logistic proliferation for the cancer cells, and production of the MMPs by both families of cancer cells. It reads itself as follows:
	\begin{equation}\label{eq:invasion}
	\left\{
	\begin{aligned}
		\frac{\partial \cc}{\partial t}
		&= D_1\nabla \cc -\chi_1\nabla\(\cc \nabla v\) + \memt c^\text{DCC} + \mu_1 \cc (1-\cc)\\
		\frac{\partial \cd}{\partial t}
		&= D_2\nabla \cd -\chi_2\nabla\(\cd \nabla v\) - \memt c^\text{DCC} + \mu_2 \cd (1-\cd)\\
		\frac{\partial v}{\partial t}
		&=-\delta m v\\
		\frac{\partial m}{\partial t}
		&=D_m\Delta m + \alpha_1\cc + \alpha_2\cd -\change{\lambda}{\beta} m
	\end{aligned}\right.
	\end{equation}
	For the rest of this work we assume that \eqref{eq:invasion} is equipped with the initial conditions:
	\begin{equation}\label{eq:ic}
	\left\{ \begin{aligned}
		\cc_0(x)& = 0,\\
		\cd_0(x)& = e^{-\frac{x^2}{0.05}},\\
		v(0,x)	& = 1- 0.5\;\cd_0(x),\\
		m(0,x)	& = 0.2\;\cd_0(x)
	\end{aligned},\quad x\in\Omega, \right .
	\end{equation}	
	where $\Omega=[0, 7.5]$.  The parameters employed are:
	\begin{align}
		\mathcal{P}=\big \{& D_1=3.8\times 10^{-4},\  D_2=3.5\times 10^{-4},\ D_m=2.5\times 10^{-3},\ \chi_1= 4\times 10^{-1},\ \chi_2= 4\times 	10^{-2},\nonumber \\
							& \mu_1=0.2,\ \mu_2=0.1,\ \alpha_1 =0.5,\ \alpha_2=0.5,\ \delta=2,\ \change{\lambda}{\beta}=1\big\}, \label{eq:params}
	\end{align}
	and for computing $\memt$, we have used 
	\begin{equation}\label{eq:params_2}
		\mathcal{P}_{\memt}= \big\{\mu_0=0.55,\  \lc=1.4,\  \ld=1,\ M = 1,\  k_D = \change{2}{0.5,\ \mu_{1/2}=2} \big\}, 
	\end{equation}
	see also \eqref{eq:muemt_2}, \eqref{explicit_K}, \eqref{eq:discr}.
	
	The above parameters in $\mathcal{P}$ were chosen in the magnitudes in which they were used in similar models (cf. \cite{Andasari.2011}, \cite{chaplain5}). Due to the properties of cancer stem cells, we chose their diffusion, and haptotaxis coefficients to be larger than the corresponding numbers for the differentiated cancer cells. The parameters in \eqref{eq:params_2} were extracted by numerical experimentation and are such that the system \eqref{eq:invasion} exhibits the current biological understanding of the problem.

	The numerical method employed on this system is a second order Finite Volume (FV) method in space combined with a third order implicit-explicit time integration scheme (IMEX3). This numerical method has been developed for such type of cancer invasion models that exhibit merging/emerging concentration phenomena. For more details we refer to our previous works \cite{Sfakianakis.2014b, Kolbe.2013} and to \cite{ck,Lukacova.2012}.

	{In} Figure \ref{fig:invasion} {we present} the dynamics of the system \eqref{eq:invasion}-\eqref{eq:params_2}. After a short period of time, some DCCs at the ECM degenerating tumor front {undergo de-differentiation} and become CSCs. They {exhibit higher motility and invasiveness than the DCCs} and escape the main body of the tumor, {they also exhibit} more dynamic phenomena than the DCCs. 

	{We further} study the influence of modifications of the EMT coefficient on the solution of the system. Therefore we compute numerical solutions of the system \eqref{eq:invasion}-\eqref{eq:params} and various $\mathcal{P}_{\memt},$ and measure the distance (in $x$) from the propagating CSC front to the slower propagating DCC front at $t=20$ as well as the total mass of CSCs and DCCs on $\Omega$ at the same time instance.

	In Figure \ref{fig:parameter_dependence} we \change{}{first }present the effect {that various values of} $k_D$ {have in the afore mentioned propagating-front-distance and CSCs, DCCs masses}. An increase of the \change{attachment/detachment}{detachment/attachment} ratio yields a decrease of the front distance and the mass of CSCs, but an increase of the total mass of DCCs. Similarly, \change{the }{}Figure \change{\ref{fig:M_dependence}}{} exhibits the {effect that $M$ has on the distances of the fronts and the masses}. Both, the front distance and the CSC mass increase with $M$ but the ascent stagnates for large $M$. The opposite is true for the mass of DCCs. \change{

	}{}\change{Finally, in Figure \ref{fig:mu_half_dependence} we present the of $\mu_{1/2}$.}{Finally, Figure \ref{fig:parameter_dependence} shows that the model proposes less DCCs and a smaller distance of the fronts for larger values of $\mu_{1/2}$.}

	\change{}{{This behavior matches our biological understanding of the EMT system. A larger amount of EGF (large $M$) strengthens EMT and thus proliferation and invasiveness of CSCs. An increase of detachment over attachment of EMT molecules ($k_D$) decreases the number of bound EGF and weakens the EMT process, which is also the case when heightening the Michaelis constant $\mu_{1/2}$. \change{If on the other hand the number of EMT molecules exceeds the number of receptors, no further catalyzing is possible. This explains the stagnation in the increase of the distance and the proliferation of the CSCs when the parameters are varied.}{The saturating behaviour at high $M$, and low $k_D$ and $\mu_{1/2}$, respectively, reflects that maximal receptor saturation and $\memt$, respectively, is reached. Thus, further increase/decrease of parameters does not change the dynamics, and distance and proliferation of CSC any more.}}}	

\section{Conclusion}

	In this work we give a mathematical insight to the EMT and \change{the subsequent ECM }{its impact on the process of cancer} invasion. In particular, we develop an algebraic-elliptic model \eqref{eq:ellip} to compute the amount of free EGF molecules in the vicinity of a tumour. We are then able to evaluate the non-constant $\memt$ coefficient that drives \change{EMT from}{the de-differentiation from} DCCs to CSCs. Subsequently, we propose a Keller-Segel type system \eqref{eq:invasion} for the ECM invasion by both types of cancer cells. 
	
	The numerical simulations that follow, although they serve here as proof of concept, exhibit clearly that the ``combined'' system is able to reproduce the current bio-medical understanding regarding the EMT first step of cancer metastasis. 
	
	Our future work includes the extension of the EMT and the ECM invasion models to 2-dimensional spaces, the inclusion of further bio-chemical interactions of the cancer cells with the environment, and the use of more biologically relevant parameters.
	
	\change{}{The effort invested in this work, and its importance is justified by the fact that in the evolution of cancer, the regulation of EMT is seen as potential drug target in cancer medicine \cite{Singh.2010,2}. A better understanding, hence, of these dynamic transitions and their relation to the cancer invasion of the ECM will be beneficial to optimize drug targeting.
	}
	

\bibliographystyle{plain} 
	\small
	\bibliography{EMT_transition}


\end{document}